\documentclass[aoas,MSNbibl,nameyear,seceqn,dvips]{arximspdf}
\usepackage{multirow,dcolumn}
\usepackage{graphicx}

\doi{10.1214/12-AOAS595} %kopijuoti is PTS
\volume{6}
\issue{4}
\pubyear{2012}
\firstpage{1588}
\lastpage{1614}

\makeatletter
\newcolumntype{d}[1]{D{.}{.}{#1}}

\newcommand{\argmin}{\operatorname{arg\,min}}
\newcommand{\argmax}{\operatorname{arg\,max}}

\def\bd{\mathbf{d}}
\def\bgamma{\bolds{\gamma}}
\def\L{\mathcal{L}}
\def\U{\mathcal{U}}
\def\cov{\operatorname{cov}}
\makeatother

\begin{document}
\begin{frontmatter}

\title{Dynamical functional prediction and classification, with
application to traffic flow prediction\thanksref{T1}}
\thankstext{T1}{Supported by Grants 99-ASIS-02 from Academia Sinica and NSC101-2118-M-001-011-MY3 from National Science Council, Taiwan.}
\runtitle{Traffic flow prediction}

\begin{aug}
\author[A]{\fnms{Jeng-Min} \snm{Chiou}\corref{}\ead[label=e1]{jmchiou@stat.sinica.edu.tw}}
\runauthor{J.-M. Chiou}
\affiliation{Academia Sinica}
\address[A]{Institute of Statistical Science\\
Academia Sinica\\
128 Sec. 2 Academia Road\\
Nankang, Taipei 11529\\
Taiwan\\
\printead{e1}} %adresu isvedimo komanda gale!
\end{aug}

% HISTORY:
\received{\smonth{11} \syear{2011}}
\revised{\smonth{8} \syear{2012}}

% ABSTRACT
%
\begin{abstract}
Motivated by the need for accurate traffic flow prediction in
transportation management, we propose a functional data method to
analyze traffic flow patterns and predict future traffic flow.
In this study we approach the problem by sampling traffic flow
trajectories from a mixture of stochastic processes.
The proposed \textit{functional mixture prediction} approach combines
functional prediction with probabilistic functional classification to
take distinct traffic flow patterns into account.
The probabilistic classification procedure, which incorporates
functional clustering and discrimination, hinges on subspace projection.
The proposed methods not only assist in predicting traffic flow
trajectories, but also identify distinct patterns in daily traffic flow
of typical temporal trends and variabilities.
The proposed methodology is widely applicable in analysis and
prediction of longitudinally recorded functional data.
\end{abstract}

% KEYWORDS
\begin{keyword}
\kwd{Clustering}
\kwd{discrimination}
\kwd{functional regression}
\kwd{mixture model}
\kwd{subspace projection}
\kwd{traffic flow rate}
\kwd{intelligent transportation system}
\end{keyword}

\end{frontmatter}

%s1 #&#
\section{Introduction}\label{sec1}

Traffic flow is an important macroscopic traffic characteristic in
transportation systems. The measurement and forecasting of traffic flow
are crucial in the design, planning and operations of highway
facilities [\citet{ZhaYe08}]. Traffic flow can be measured
automatically using various types of vehicle detectors such as the
commonly used dual loop detectors, which are installed in certain roads
at regular intervals. Real-time traffic flow information in conjunction
with historical traffic flow records makes it possible to predict
traffic flow in the short term. The importance of traffic prediction
for intelligent transportation systems has long been recognized in many
applications, including the development of traffic control strategies
in advanced traffic management systems and real-time route guidance in
advanced traveler information systems [\citet{ZheLeeShi06}].
However, dynamic features of traffic flow, along with unstable traffic
conditions and unpredictable environmental factors, contribute to the
challenge of pursuing accuracy in predictions.

Short-term traffic flow prediction has been intensively investigated
for more than two decades and various types of methodologies have been
developed.
These include time series models [e.g., \citet{WilHoe03},
\citet{StaKar03}], Kalman filtering methods [e.g.,
\citet{XieZhaYe07}, \citet{OkuSte84}],
local linear regression models [\citet{Sunetal03}],
neural network based methods [e.g., \citet{CheGra01},
\citet{ZheLeeShi06}, \c{C}etiner, Sari and Borat (\citeyear
{CetSarBor10})]
and
fuzzy neural models and fuzzy logic system methods
[\citet{Yinetal02},
\citet{ZhaYe08}], among others.
In addition, there are many articles comparing parametric time series
models, nonparametric regression models and neural networks in traffic
prediction, such as in \citet{KirWasDou97}, \citet
{SmiDem97} and \citet{SmiWilOsw02}.
More recently, \citet{KamSheWyn12} discussed road
traffic forecasting for highway networks using
fully parametric regime-switching space--time models, coupled with a
penalized estimation scheme.
To our knowledge, a functional data approach to predicting traffic flow
has not yet been investigated in the literature.

%
%f1 #&#
\begin{figure}[b]

\includegraphics{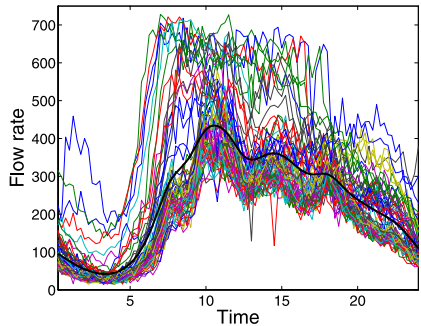}

\caption{Daily traffic flow trajectories (training data) with the
estimated mean function superimposed on the observed trajectories.}
\label{figFlow-train-mu}
\end{figure}

%s1.1 #&#
\subsection{Illustration of traffic flow prediction and the proposed
functional data method} \label{subsectraffic}
\begin{figure}

\includegraphics{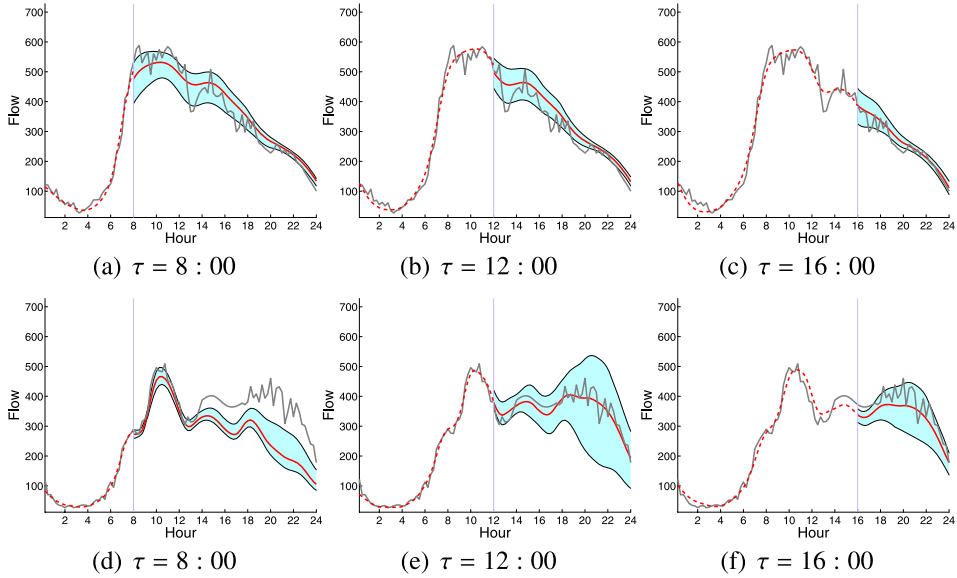}

\caption{Two samples of trajectories from the test data set
[\textup{(a)--(c)} for Test sample 1; \textup{(d)--(f)} for Test
sample 2]. The fitted
curves (dotted lines before times $\tau$) and the predicted curves
(solid curves after times $\tau$) with 95\% prediction intervals for a
partially observed trajectory available up to times $\tau=8$, 12 and
16:00, superimposed on the complete trajectory (gray line).}
\label{figFlowctx0cty0-sub3}
\end{figure}

Motivated by a practical need for accurate traffic flow prediction, we
develop a novel functional data method for predicting future, or
unobserved, daily traffic flow for an up-to-date and partially observed
traffic flow trajectory. Figure~\ref{figFlow-train-mu} illustrates a
sample of daily traffic flow trajectories. The data were collected by a
dual loop vehicle detector located near Shea-San Tunnel on National
Highway 5 in Taiwan in 2009 and are based on the flow rates (vehicle
count per min) over 15-min time intervals, a metric suggested in
\textit
{Highway Capacity Manual 2000} for operational analyses [\citet
{ZheLeeShi06}]. The trajectories sample 70 days as the training data,
while the remaining 14 days are used as the test data to validate the
prediction performance. The aim is to predict the unobserved traffic
flow trajectory for a partial trajectory with updated flow information
up to the ``current time'' $\tau$, which is given as a time of day.
In Figure~\ref{figFlowctx0cty0-sub3}, the raw trajectories (gray
lines) before $\tau=8$, 12 and 16:00 are observations from the test
data, superimposed on the curves (dotted lines) fitted by functional
principal component analysis.
After the last observation time point $\tau$, the predicted traffic
flow trajectories (solid lines) are obtained by the proposed functional
mixture prediction model coupled with the 95\% bootstrap prediction intervals.
The real data trajectories after times $\tau$ (gray lines) are
unobserved in the prediction scenario and are displayed for comparative
purposes.
The prediction for the trajectory is dynamically updated as the
``current time'' $\tau$ progresses.

%
%f2 #&#

% \begin{center} \scriptsize
% \subfigure[$\tau=8:00$]{
% \includegraphics[scale=0.21]{Fig2a-Flow_ctx0_cty0-sub3_CT_1.eps}
% \label{figFlowCT8ctx0cty0-sub3}
% }
% \subfigure[$\tau=12:00$]{
% \includegraphics[scale=0.21]{Fig2b-Flow_ctx0_cty0-sub3_CT_17.eps}
% \label{figFlowCT12ctx0cty0-sub3}
% }
% \subfigure[$\tau=16:00$]{
% \includegraphics[scale=0.21]{Fig2c-Flow_ctx0_cty0-sub3_CT_33.eps}
% \label{figFlowCT16ctx0cty0-sub3}
% }
% \subfigure[$\tau=8:00$]{
% \includegraphics[scale=0.21]{Fig2d-Flow_ctx0_cty0-sub1_CT_1.eps}
% \label{figFlowCT8ctx0cty0-sub1}
% }
% \subfigure[$\tau=12:00$]{
% \includegraphics[scale=0.21]{Fig2e-Flow_ctx0_cty0-sub1_CT_17.eps}
% \label{figFlowCT12ctx0cty0-sub1}
% }
% \subfigure[$\tau=16:00$]{
% \includegraphics[scale=0.21]{Fig2f-Flow_ctx0_cty0-sub1_CT_33.eps}
% \label{figFlowCT16ctx0cty0-sub1}
% }
%
% \end{center}

We note that the aforementioned methods in the literature were largely
developed based on ``short-term'' or ``next-step'' traffic prediction
modeling, where ``short-term'' refers to a forecast horizon of a time interval.
A 15-min interval is commonly used as the forecast horizon for
traveler-oriented applications and operational analysis [\citet{ZhaYe08}].
In contrast to the ``next-interval'' prediction methods in the existing
literature, our functional data method is more flexible, as illustrated
in Figure~\ref{figFlowctx0cty0-sub3}, allowing valid prediction
periods extended from the ``current time'' to the end of the day and
thus providing more information to relevant users.

Since future traffic conditions and temporal traffic flow patterns play
a critical role in traffic prediction [e.g., \citet{SmiWilOsw02},
\citet{VlaKarGol08}], to take into account
distinct daily traffic flow patterns,
we propose a \textit{functional mixture prediction} approach that
combines functional prediction with a probabilistic classification procedure.
Specifically, we propose to implement functional cluster analysis of
past traffic flow trajectories to obtain typical daily traffic flow
patterns or clusters, followed by a probabilistic classification for
the traffic flow trajectory observed thus far. Based on the traffic
flow patterns or clusters identified by the proposed method, we can
predict the
unobserved traffic flow trajectories by a functional prediction model
in conjunction with the estimated posterior membership probabilities of
traffic flow clusters.
Although motivated by traffic flow analysis and prediction,
the proposed methodology is by no means restricted to this particular
field and is generally applicable to a wide variety of longitudinally
recorded functional data.

In many real applications, clustering of curve data can be challenging
and misclassification of an up-to-date and partially observed
trajectory can cause loss of prediction accuracy. Hence, a simple
``prediction-after-classification'' approach based on hard clustering
results may not be the best approach. This will be illustrated in our
numerical studies including the real data application and simulations.
In contrast, the proposed functional mixture prediction framework
addresses challenges related to prediction of complex functional data,
such as those containing heterogeneous patterns and large variability
over time.
The proposed \textit{functional mixture prediction} approach to traffic
flow prediction has the following features:
\begin{itemize}
\item It is the first approach to employ functional data techniques to
address traffic flow prediction applications, which are critical in
many intelligent transportation systems.
\item The functional data approach allows for interval prediction in
contrast to ``next-step'' traffic flow prediction applications found in
the literature.
\item The proposed functional mixture prediction model plays a central
role in predicting the future trajectory for an up-to-date and
partially observed trajectory and takes distinct traffic flow patterns
into account to improve prediction accuracy.
This study extends the idea of the subspace projected functional
clustering method of \citet{ChiLi07} to identify distinct patterns
of daily traffic flow from the past data, coupled with the forward
functional testing procedure of \citet{LiChi11} to determine the
number of clusters, which lays the groundwork for the functional
mixture prediction approach.
\item The probabilistic classification approach, including functional
clustering and discrimination, is new. It
allows for the prediction of posterior membership probabilities using
partial information up to the current time, in contrast to clusters
constructed by complete trajectories from past data.
\item The predictive function in the functional mixture prediction
model is built on an existing functional linear model, which is widely
used in functional regression modeling [e.g., \citet{RamDal91},
M\"uller, Chiou and Leng (\citeyear{Mul09})] and is easy to implement.
\end{itemize}

%s1.2 #&#
\subsection{Literature review of relevant functional data methods}\label{sec1,2}

Statistical tools for functional data analysis have been extensively
developed during the past two decades to deal with data samples
consisting of curves or other infinite-dimensional data objects.
Systematic overviews of functional data analysis are provided in the
monographs of Ramsay and Silverman (\citeyear{RamSil02}, \citeyear
{RamSil05}) and \citet{FerVie06} and in the review articles of
\citet{Ric04}, \citet{ZhaMarWel04}
and M\"uller (\citeyear{Mul05,Mul09}). Functional data analysis
provides a wide range of applications in many disciplines.
These include biomedical and environmental studies [\citet
{Dietal09}, \citet{GaoNie08}], analysis of time-course gene
expression profiles
[M\"uller, Chiou and Leng (\citeyear{MulChiLen08}), \citet
{CofHin11}], linguistic
pitch analysis [\citet{AstChiEva10}] and demographic and
mortality forecasting [Hyndman and Shahid Ullah (\citeyear{HynSha07}),
Chiou and M\"uller
(\citeyear{ChiMul09}), \citet{DAmPisRus11}], among many others.
In relation to functional data prediction, M\"uller and Zhang
(\citeyear{MulZha05})
proposed a functional data approach to predicting remaining lifetime
and age-at-death distributions from individual event histories observed
up to the current time.
More recently, \citet{ZhoSerGeb11} proposed a functional
data approach to degradation modeling for the evolution of degradation
signals and the remaining life distribution. These are relevant works
that contain novel functional data techniques with interesting
applications to the prediction of an unobserved event for a partial
trajectory observed up to the current time.

Among the various settings in functional regression analysis [M\"uller
(\citeyear{Mul05})], models with both the response and predictor
variables as
functions serve this study's purpose with regard to prediction.
Functional regression models of this kind have been considered, for
example, in Yao, M\"uller and Wang (\citeyear{YaoMulWan05N2}), Chiou
and M\"uller (\citeyear{ChiMul07}),
M\"uller, Chiou and Leng (\citeyear{MulChiLen08}) and \citet{Antetal10}.
Methods of functional data clustering that are found in the literature
include the use of multivariate clustering algorithms on the
finite-dimensional coefficients of basis function expansions [e.g.,
\citet{Abretal03}, \citet{SerWas05}], model-based
functional data clustering [e.g., \citet{JamSug03}, \citet
{MaZho08}], a general descending hierarchical algorithm [Chapter 9 of
\citet{FerVie06}] and various depth-based classification methods
[\citet{CueFebFra07}, L\'opez-Pintado and Romo (\citeyear
{LopRom06})], among others.
Of particular interest with regard to functional prediction models are
the methods that define clusters via subspace projection [Chiou and Li
(\citeyear{ChiLi07}, \citeyear{ChiLi08})].
The subspace projection method considers cluster differences not only
in mean functions, but also in eigenfunctions of covariance kernels
that takes into account individual random process variations, making it
suitable for interpreting the stochastic nature of traffic flow and
suggesting a natural link with functional regression models.

This article is organized as follows. In Section \ref{clustering&classification} we represent traffic
flow trajectories as a mixture of stochastic processes and discuss
functional clustering and classification methods to take traffic flow
patterns into account.
Section \ref{sec3} discusses the functional mixture prediction model, including
the algorithm for implementing functional mixture prediction.
Sections \ref{sec4} and \ref{sec5} illustrate the empirical analysis of traffic flow
patterns and results of predicting traffic flow trajectories.
Section~\ref{sec6} presents a simulation study to evaluate the performance of
the functional mixture prediction in comparison with related methods.
Concluding remarks and discussion are provided in Section~\ref{sec7}.
More information in selecting the number of clusters, the bootstrap
prediction intervals and additional details in the simulation design
and results are deferred to Supplementary Materials [\citet{chi12}].

%s2 #&#
\section{Modeling traffic flow trajectories and clustering traffic flow
patterns} \label{clustering&classification}

Previous studies in traffic flow prediction and modeling have revealed
that traffic condition data is characteristically stochastic, as
opposed to chaotic [\citet{SmiWilOsw02}].
The stochastic features of traffic flow trajectories are suggestive of
a functional data approach.
In the functional data framework, we adopt the notion that
each daily traffic flow trajectory is a realization of a random function
sampled from a mixture of stochastic processes.
Let $Z$ denote the random function for the daily traffic flow
trajectory in the domain $\U=[0,T]$. Here, the random function $Z$ is
square integrable
with the inner product of any two functions $f_1$ and $f_2$
defined as $\langle f_1, f_2 \rangle_{\U} = \int_{\U} f_1(t) f_2(t) \,dt$
with the norm $\|f_1\|_{\U}=\langle f_1, f_1 \rangle_{\U}^{{1}/{2}}$.
It is assumed that the random function $Z(t)$ has a smooth mean
function $E Z(t)=\mu_Z(t)$ and covariance function $\cov
(Z(s),Z(t))=G_Z(s,t)$, for $s$ and $t$ in $\U$.

%s2.1 #&#
\subsection{Functional clustering of historical traffic flow
trajectories} \label{clustering}

While temporal traffic flow patterns are critical in traffic
prediction, the underlying traffic flow structures and number of
typical patterns are unknown and remain to be explored.
We assume the mixture process $Z$ consists of $K$ subkprocesses, with
each subprocess corresponding to a cluster.
The random cluster variable $C$ for each individual cluster membership
is randomly distributed among clusters with label $c\in\{1,\ldots, K\}$.
For each subprocess associated with cluster $c$, define the conditional
mean function $E(Z(t)\mid C=c)=\mu^{(c)}(t)$ and covariance function
$\cov(Z(s),Z(t) \mid C=c)=G^{(c)}(s,t)$, for $c\in\{1,\ldots,K\}$.
Let $(\lambda^{(c)}_j, \varphi^{(c)}_j)$ be the corresponding
eigenvalue--eigenfunction pairs of the covariance kernel $G^{(c)}$,
where $\lambda^{(c)}_j$ are in nonascending order.
Assume, under mild conditions, each subprocess possesses a Karhunen--Lo\`
{e}ve expansion
for the daily traffic flow trajectory $Z$ given by
%
%
%e2.1 #&#
\begin{equation}
\label{KLZc} {Z}^{(c)}(t) = \mu^{(c)}(t) + \sum
_{j = 1}^{\infty} {\xi}^{(c)}_{j}
\varphi^{(c)}_{j}(t) ,
\end{equation}
where ${\xi}^{(c)}_{j} = \langle Z-\mu^{(c)},\varphi^{(c)}_j\rangle_{\U
}$ with $\langle\varphi^{(c)}_j, \varphi^{(c)}_l\rangle_{\U}=1$ for
$j=l$ and 0 otherwise.
In practice, it is often the case that the representation only requires
a small number of components to approximate the trajectories.
In general, trajectories with simpler structure require fewer
components as compared to more complex trajectories.

Following the conventional approach, the best cluster membership $c^*$
given $Z$ is determined by maximizing the posterior probability
$P_{C\mid Z}(\cdot\mid\cdot)$ such that
%
%
%e2.2 #&#
\begin{equation}
\label{ccondprob} c^*(Z)=\mathop{\argmax}_{c\in\{1, \ldots, K\}} P_{C \mid Z}(c\mid Z).
\end{equation}
We propose estimating the posterior membership probability $P(C=c \mid
Z)$ using the so-called discriminative approach, as opposed to the
generative approach [see, e.g., \citet{Daw76}, Bishop and
Lasserre (\citeyear{BisLas07})].
While there is no general consensus for choosing between generative and
discriminative approaches [\citet{NgJor02}, \citet
{XueTit08}], the former
requires {a priori} knowledge on the class-conditional
probability density functions, information that is difficult to justify
incorporating for the traffic flow trajectories. It is easier to use
the discriminative approach that directly estimates the
class-membership probabilities without attempting to model the
underlying probability distributions of the random functions.
Following this line, the multiclass logit model is a popular method for
estimating the posterior membership probabilities. We propose
incorporating a distance measure between $Z$ and its projection
associated with each cluster as the covariate in the multiclass logit model.

Consider the relative $L^2$ distance as the distance measure based on
cluster subspace projection as
%
%
%e2.3 #&#
\begin{equation}
\label{drelative} d^{(c)}=\frac{\| Z - \tilde{Z}^{(c)} \|^2}{\sum_{k=1}^K \| Z -
\tilde
{Z}^{(k)} \|^2} ,
\end{equation}
where $\tilde{Z}^{(c)}(t) = \mu^{(c)}(t) + \sum_{l = 1}^{M_c} {\xi
}_{l}^{(c)} \varphi_{l}^{(c)}(t)$, with ${\xi}_{l}^{(c)} = \langle
Z-\mu^{(c)}, \varphi^{(c)}_{l}\rangle_{\U}$.
The value $M_c$ is finite and is chosen data-adaptively so that $Z$ is
well approximated by $Z^{(c)}$ by the $M_c$ components.
Let $\bd=(1, d^{(1)}, \ldots, d^{(K-1)})^\top$ and $\bgamma_c=(\gamma_{0c}, \gamma_{1c}, \ldots, \gamma_{(K-1)c})^\top$.
Taking the vector $\bd$ as the covariate, we can estimate the posterior
cluster membership probability using the multiclass logit model,
%
%e2.4 #&#
\begin{eqnarray}
\label{mclogitK} P(C=c \mid Z) = \frac{\exp\{ \bgamma_c^\top\bd\}}{ \sum_{k=1}^K
\exp
\{ \bgamma_k^\top\bd\}}
\end{eqnarray}
for $c=1, \ldots, K-1$ and $P(C=K \mid Z)=1- \sum_{c=1}^{K-1} P(C=c
\mid Z)$ with the $K$th cluster being the baseline.
The vector of regression coefficients $\bgamma_{c}$ remains to be estimated.

Clusters defined by criterion~(\ref{ccondprob}) are based on subspace
projection.
Let $S^{(c)}_{M}$ be the linear span of the set of eigenfunctions $\{
\varphi_1^{(c)}, \ldots, \varphi_{M_c}^{(c)}\}$, $c=1,\ldots,K$.
For identifiability, it is assumed that for any two clusters $c$ and
$d$ the following two conditions do not hold simultaneously: (i)
$S^{(c)}_{M}$ belongs to $S^{(d)}_{M}$, (ii) $\mu^{(c)}=\mu^{(d)}$, or
$\mu^{(c)}\in S^{(d)}_{M}$ and $\mu^{(d)}\in S^{(c)}_{M}$. These
conditions were derived in Theorem 1 of \citet{ChiLi07} for
identifiability of clusters defined via subspace projection.
Criterion~(\ref{ccondprob}) leads to clusters with similar curves
that are embedded in the cluster subspace spanned by the cluster center
components, the mean function and the eigenfunctions of the covariance
kernel that represent the functional principal component subspace.

In addition, the number of clusters is unknown, and must be determined
in practice. The method used to determine the number of clusters in
this study is based on a sequence of tests on cluster structures done
to ensure statistical significance in the difference between cluster
types as proposed in \citet{LiChi11}.
The number of clusters $K$ is determined by testing a sequence of null
hypotheses
$H_{01}\dvtx \mu^{(c)} = \mu^{(d)}$ and $H_{02}\dvtx S^{(c)}_{M} = S^{(d)}_{M}$,
for $1\le c\ne d\le K$.
The \textit{forward functional testing} procedure aims to search for
the maximum number of clusters while retaining differences with
statistical significance among the clusters. The procedure is
especially suitable for the subspace projected functional clustering
method. The sequence of the functional hypothesis tests helps identify
significant differences between cluster structures and
provides additional insight into further cluster analysis.
Since the hypothesis tests are based on bootstrap resampling methods,
it takes substantial computational time to construct the reference distribution.
Details of the procedure are discussed in \citet{LiChi11} and we
have briefly summarized them in Supplementary Material A [\citet{chi12}].

%s2.2 #&#
\subsection{Probabilistic functional classification of traffic flow
patterns} \label{classification}

For the purpose of prediction, the time domain $\U$ of the process $Z$
is decomposed into two exclusive time domains ${\mathcal{S}(\tau
)}=[0,\tau]$ and ${\mathcal{T}(\tau)}
=[\tau, T]$.
Now, let $Z^*$ be a newly observed trajectory of the process $Z$,
denoted by $Z_{\mathcal{S}(\tau)}^*$ as observed up to time $\tau$.
We predict the cluster membership probability of the trajectory $Z^*$
based on the known trajectory $Z_{{\mathcal{S}(\tau)}}^*$ observed
until time $\tau$,
which will then be used to predict the unobserved trajectory
$Z_{{\mathcal{T}(\tau)}}^*$.

We define the relative $L^2$ distance in a manner similar to~(\ref
{drelative}) via cluster subspace projection, but it is based on the
partially observed $Z_{{\mathcal{S}(\tau)}}^*$ rather than the entire
$Z^*$ since the
part $Z_{{\mathcal{T}(\tau)}}^*$ is not yet observed. Suppose that
the cluster subspaces
$\mu^{(c)}$ and $\{\varphi_{{\mathcal{S}(\tau)},j}^{(c)}\}$, $c=1,
\ldots, K$, are being
identified as in Section~\ref{clustering}. Then, the relative $L^2$
distance is defined as
%
%
%e2.5 #&#
\begin{equation}
\label{dS} d_{\mathcal{S}(\tau)}^{*(c)}=\frac{\| Z_{{\mathcal{S}(\tau)}}^* -
\tilde{Z}_{\mathcal{S}(\tau)}^{*(c)} \|^2}{\sum_{k=1}^K
\| Z_{{\mathcal{S}(\tau)}}^* - \tilde{Z}_{\mathcal{S}(\tau)}^{(k)}
\|^2} ,
\end{equation}
where $\tilde{Z}_{\mathcal{S}(\tau)}^{*(c)}(s) = \mu^{(c)}(s) +
\sum_{l = 1}^{M_c}
{\xi
}_{{\mathcal{S}(\tau)},l}^{*(c)} \varphi_{{\mathcal{S}(\tau
)},l}^{(c)}(s)$, with ${\xi}_{{\mathcal{S}(\tau)}
,l}^{*(c)} =
\langle Z_{\mathcal{S}(\tau)}^*- \mu^{(c)},\break \varphi^{(c)}_{{\mathcal
{S}(\tau)},l}\rangle_{{\mathcal{S}(\tau)}}$.
Here, the set of eigenfunctions $\{\varphi^{(c)}_{{\mathcal{S}(\tau
)},l}\}$ corresponds
to the covariance kernel $G^{(c)}_{{\mathcal{S}(\tau)}}$ of the
random process $Z_{{\mathcal{S}(\tau)}}$.
Taking $\bd_{\mathcal{S}(\tau)}^*=(1, d_{\mathcal{S}(\tau
)}^{*(1)}, \ldots,\break  d_{\mathcal{S}(\tau)}^{*(K-1)})^\top$ as the
covariate, we can predict the cluster membership probability based on
the newly observed $Z^*_{\mathcal{S}(\tau)}$ using the multiclass
logit model
%
%e2.6 #&#
\begin{equation}
\label{mclogitK*} P\bigl(C=c \mid Z_{\mathcal{S}(\tau)}^*\bigr) = \frac{\exp\{ \bgamma_c^\top
\bd_{\mathcal{S}(\tau)}^* \}}{
\sum_{k=1}^K \exp\{ \bgamma_k^\top\bd_{\mathcal{S}(\tau)}^* \}}
\end{equation}
for $c=1, \ldots, K-1$, and $P(C=K \mid Z_{\mathcal{S}(\tau)}^*)=1-
\sum_{c=1}^{K-1}
P(C=c \mid Z_{\mathcal{S}(\tau)}^*)$ with the $K$th cluster being baseline.
We note that the vector of coefficients $\bgamma_c$ here is the same as
that in
(\ref{mclogitK}) based on the historical or training data.

%s2.3 #&#
\subsection{Estimation for probabilistic functional classification}
\label{estimationclustering}

In practice, the observed trajectories may be contaminated with random
measurement errors. Let $Y_i(t_{ij})$ be the $j$th observation of the
$i$th individual flow trajectory from the underlying process
$Z_i^{(c)}$ of cluster $c$ observed at time $t_{ij}$, $0\le t_{ij}\le
T$, such that
$Y_i(t_{ij}) = Z_i^{(c)}(t_{ij}) + \epsilon_{ij}$,
where $Z_i^{(c)}$ is the underlying random function such that
$Z_i^{(c)}(t)=\mu^{(c)}(t) + \sum_{k = 1}^{\infty}{\xi}_{ki}^{(c)}
\varphi_{k}^{(c)}(t)$, and the random measurement errors $\epsilon_{ij}$ are independent of ${\xi}_{ki}^{(c)}$ with mean zero and
variance $\sigma^2$.

To identify the structures of cluster subspaces, $\{\hat\mu^{(c)}, \{
\hat\varphi_k^{(c)}\}_{k=1,\ldots,M_c,}\}$, $c=1, \ldots,K$, we follow
the idea of defining clusters via subspace projection and apply the
proposed subspace-projected functional clustering procedure to the
training data set.
In the initial step, since cluster membership is unknown, the
clustering is based on functional principal component scores of an
overall single random process. Details of the initial clustering refer
to Section 2.2.1 of \citet{ChiLi07}. In the iterative updating
step, cluster membership is determined by criterion~(\ref{ccondprob})
in a hard clustering manner. The clustering procedure is implemented
iteratively in identifying between (i) cluster subspaces and (ii)
cluster memberships until convergence.

\subsubsection*{Cluster subspaces} Given the observations $\{
(t_{ij},Y_i(t_{ij})), i=1, \ldots, n, j=1, \ldots, m\}$, from the
historical or training data, and the cluster memberships of the
trajectories, using the observations belonging to cluster $c$, the mean
function $\mu^{(c)}$ can be estimated by applying the locally weighted
least squares method while the estimates of the components
$\varphi^{(c)}_{k}$ and ${\xi}^{(c)}_{ik}$ rely on the covariance
estimate $\hat{G}^{(c)}$ by applying the smoothing scatterplot data
$(Y_{ij}-\hat\mu^{(c)}(t_{ij}))(Y_{il}-\hat \mu^{(c)}(t_{il}))$ to fit
a local linear plane. Details of this estimation can be found in Chiou,
M\"uller and Wang (\citeyear {ChiMulWan03}) and Yao, M\"uller and Wang
(\citeyear{YaoMulWan05N1}), for example. The smoothing parameters in
the mean and covariance estimation steps are chosen data-adaptively via
the 10-fold cross-validation method. An estimate of ${\xi}^{(c)}_{ik}$
can be obtained by the conditional expectation approach of
\citet{YaoMulWan05N1} for the case of sparse designs. Here, we
simply obtain the estimate $\hat{\xi}^{(c)}_{ik}=\langle
Y_i-\hat\mu^{(c)}, \hat\varphi_{k}{(c)}\rangle_{\U}$ by numerical
approximation for the case of dense designs of traffic flow recording.
The value $M_c$ is selected as the minimum that reaches a certain level
of the proportion of total variance explained by the $M_c$ leading
components such that
%
%
%e2.7 #&#
\begin{equation}
\label{Mc} M_c = \mathop{\argmin}_{L: L\ge1} \Biggl\{ \sum
_{l=1}^L \hat\lambda_l^{(c)}
\bigg/ \sum_{l=1}^\infty\hat
\lambda_l^{(c)} 1_{\{\hat\lambda_l^{(c)}>0\}} \ge\delta \Biggr\},
\end{equation}
where $\delta=90\%$ in this study.
These cluster structure estimates are used in turn to estimate the
vector of regression coefficients $\bgamma_k$ in~(\ref
{mclogitKhat}) below.

\subsubsection*{Cluster memberships} Given the structure of each cluster
based on the training data, we then use the discriminative approach to
fit the posterior probabilities of cluster membership $P(C=c \mid Y_i)$
such that
%
%e2.8 #&#
\begin{eqnarray}
\label{mclogitKhat} \hat{P}(C=c \mid Y_i) = \frac{\exp\{\hat{\bgamma}_c^\top\hat
{\bd
}_i\}}{ \sum_{k=1}^K
\exp\{\hat{\bgamma}_k^\top\hat{\bd}_i\}} ,\qquad  c=1,
\ldots, K-1,
\end{eqnarray}
and $\hat{P}(C=K \mid Y_i)=1- \sum_{k=1}^{K-1} \hat{P}(C=k \mid Y_i)$,
taking the $K$th cluster as the baseline. Here, the relative $L^2$
distance vector $\hat\bd_i=(1, \hat{d}_{i}^{(1)}, \ldots, \hat
{d}_{i}^{(K-1)})^\top$ serves as the predictor variable and is
calculated by
%
%
%e2.9 #&#
\begin{equation}
\label{dSY} \hat{d}_{i}^{(c)}= \frac{\| Y_i - \hat{Z}_i^{(c)} \|^2} {\sum_{k=1}^K \| Y_i - \hat
{Z}_i^{(k)} \|^2} ,\qquad c=1,
\ldots, K-1,
\end{equation}
where
$\hat{Z}_i^{(c)}(s) = \hat{\mu}^{(c)}(s) + \sum_{j = 1}^{M_c} \hat
{\xi
}_{ij}^{(c)} \hat{\varphi}_{j}^{(c)}(s)$, with $\hat{\xi}^{(c)}_{ij}$
defined above.
The coefficient estimates $\hat{\bgamma}_c=(\gamma_{0c}, \gamma_{1c},
\ldots, \gamma_{(K-1)c})$ are obtained by the conventional iterated
reweighted least squares method [McCullagh and Nelder (\citeyear{McCNel83})].
The resulting estimate~(\ref{mclogitKhat}) is used to determine the
cluster membership according to~(\ref{ccondprob}).

Now, given a newly observed trajectory $Y^*$ up to time $\tau$, denoted
by $Y^*_{{\mathcal{S}(\tau)}}$, we obtain the covariate vector $\hat
\bd_{\mathcal{S}(\tau)}^*=(1,
\hat
{d}_{\mathcal{S}(\tau)}^{*(1)}, \ldots, \hat{d}_{\mathcal{S}(\tau
)}^{*(K-1)})^\top$, where
%
%
%e2.10 #&#
\begin{equation}
\label{dSY*} \hat{d}_{\mathcal{S}(\tau)}^{*(c)}= \frac{\| Y_{{\mathcal{S}(\tau
)}}^* - \hat{Z}_{{\mathcal{S}(\tau)}}^{*(c)} \|^2}{\sum_{k=1}^K \| Y_{{\mathcal{S}(\tau)}}^* - \hat{Z}_{{\mathcal{S}(\tau
)}}^{(k)} \|^2} , \qquad c=1,
\ldots, K-1.
\end{equation}
Here,
$\hat{Z}_{{\mathcal{S}(\tau)}}^{*(c)}(s) = \hat\mu^{(c)}(s) + \sum_{j = 1}^{M_c}
\hat{\xi
}_{{\mathcal{S}(\tau)},j}^{*(c)} \hat\varphi_{{\mathcal{S}(\tau
)},j}^{(c)}(s)$,
and $\hat{\xi}^{*(c)}_{{\mathcal{S}(\tau)},j}$ can be obtained by a
numerical
approximation to
$\langle Z^*_{{\mathcal{S}(\tau)}}-\hat\mu_{{\mathcal{S}(\tau
)}}^{(c)}, \hat\varphi_{{\mathcal{S}(\tau)}
,j}^{(c)}\rangle_{{\mathcal{S}(\tau)}}$.
To obtain\break $\{\hat\varphi_{{\mathcal{S}(\tau)},j}^{(c)}\}$, we
simply decompose the
covariance estimate $\hat{C}^{(c)}$ into blocks corresponding to the
time domains ${\mathcal{S}(\tau)}$ and ${\mathcal{T}(\tau)}$
without re-estimating the covariance
function, making the dynamical prediction step easy to implement for
any given $\tau$.
We predict the cluster membership for the newly observed $Y^*_{\mathcal
{S}(\tau)}$ by
the posterior probability
%
%e2.11 #&#
\begin{eqnarray}
\label{mclogitK*hat}
\quad \hat P\bigl(C=c \mid Y_{{\mathcal{S}(\tau)}}^*\bigr) =
\frac{\exp\{ \hat
{\bgamma}_c^\top
\hat\bd_{\mathcal{S}(\tau)}^* \}}{ \sum_{k=1}^K
\exp\{ \hat{\bgamma}_k^\top\hat\bd_{\mathcal{S}(\tau)}^* \}} ,\qquad c=1, \ldots , K-1,
\end{eqnarray}
and $\hat P(C=K \mid Y_{{\mathcal{S}(\tau)}}^*)=1- \sum_{c=1}^{K-1}
\hat P(C=c \mid
Y_{{\mathcal{S}(\tau)}}^*)$.

%s3 #&#
\section{Functional mixture prediction of future traffic flow trajectories}\label{sec3}

To accurately predict traffic flow trajectories under various traffic
conditions, we combine the functional prediction model with functional
clustering and classification methods.
Given a newly observed trajectory $Z_{\mathcal{S}(\tau)}^*$ of the
process $Z$ as
observed up to time $\tau$, we propose a \textit{functional mixture
prediction model} to predict the trajectory of $Z^*$ on the time
interval ${\mathcal{T}(\tau)}=[\tau,T]$, denoted by $Z_{{\mathcal
{T}(\tau)}}^*$ as
%
%
%e3.1 #&#
\begin{equation}
\label{mixturepred} E\bigl(Z_{{\mathcal{T}(\tau)}}^*(t) \mid Z_{{\mathcal{S}(\tau)}}^*\bigr) =
\sum_{c=1}^K P\bigl(C=c \mid
Z_{{\mathcal{S}(\tau)}}^*\bigr) \tilde {Z}_{{\mathcal{T}(\tau)}}^{*(c)}(t) ,
\end{equation}
where $\tilde{Z}_{{\mathcal{T}(\tau)}}^{*(c)}(t)= E(Z^*_{{\mathcal
{T}(\tau)}}(t) \mid Z_{{\mathcal{S}(\tau)}}^*, C=c)$ is
the predictive function conditional on cluster $C=c$, and $P(C=c \mid
Z_{{\mathcal{S}(\tau)}}^*)$ is the posterior probability of cluster
membership given the
newly observed trajectory $Z_{{\mathcal{S}(\tau)}}^*$ up to time
$\tau$.
The functional mixture prediction model~(\ref{mixturepred}), obtained
by the law of iterated expectation on the random cluster membership
variable $C$, $E \{E(Z_{{\mathcal{T}(\tau)}}^*(t) \mid Z_{{\mathcal
{S}(\tau)}}^*, C)\}$,
minimizes the expected risk,
$E  \{\L(Z_{{\mathcal{T}(\tau)}}^*,\widetilde{Z}_{{\mathcal
{T}(\tau)}}^C)  \}$, where
$\widetilde
{Z}_{{\mathcal{T}(\tau)}}^C(t)= E(Z^*_{{\mathcal{T}(\tau)}}(t) \mid
Z_{{\mathcal{S}(\tau)}}^*, C)$ and the loss function
is defined as $\L(Z_{{\mathcal{T}(\tau)}}^*,\widetilde
{Z}_{{\mathcal{T}(\tau)}}^c)=\int_{{\mathcal{T}(\tau)}}
\{ Z_{{\mathcal{T}(\tau)}
}^*(t) - \tilde{Z}_{{\mathcal{T}(\tau)}}^{*(c)}(t)  \}^2 \,dt$.

%s3.1 #&#
\subsection{Functional linear regression of traffic flow trajectories}
\label{FLR}

In a regression setting, the process $Z(s)$, for $s\in{\mathcal
{S}(\tau)}$ denoted by
$Z_{\mathcal{S}(\tau)}$, serves as the predictor function and the
process $Z(t)$, for
$t\in{\mathcal{T}(\tau)}$ denoted by $Z_{\mathcal{S}(\tau)}$, is
the response function.
The subspace projected functional clustering method developed above is
well suited to identifying clusters in conjunction with functional prediction.
For each cluster subspace, $Z^{(c)}$ is decomposed into $Z_{{\mathcal
{S}(\tau)}
}^{(c)}(s)$ and $Z_{{\mathcal{T}(\tau)}}^{(c)}(t)$ whose Karhunen--Lo\`
eve expansions can
be obtained such that $Z_{\mathcal{S}(\tau)}^{(c)}(s)= \mu^{(c)}(s)
+ \sum_{j=1}^{\infty
} \xi_{{\mathcal{S}(\tau)},j}^{(c)} \varphi_{{\mathcal{S}(\tau
)},j}^{(c)}(s)$ and $Z_{\mathcal{T}(\tau)}^{(c)}(t)=
\mu^{(c)}(t) + \sum_{j=1}^{\infty} \xi_{{\mathcal{T}(\tau)},j}^{(c)}
\varphi_{{\mathcal{T}(\tau)}
,j}^{(c)}(t)$, where the notation $\xi_{{\mathcal{S}(\tau
)},j}^{(c)}$, $\varphi_{{\mathcal{S}(\tau)}
,j}^{(c)}$, $\xi_{{\mathcal{T}(\tau)},j}^{(c)}$ and $\varphi_{{\mathcal{T}(\tau)},j}^{(c)}$ are defined
analogously to those on the entire domain $\U$, but they correspond to
the sub-domains ${\mathcal{S}(\tau)}$ and ${\mathcal{T}(\tau)}$.

We consider a functional linear regression model [e.g., \citet
{RamDal91}, M\"uller, Chiou and Leng (\citeyear{MulChiLen08})]
conditional on cluster
membership,
%
%
%e3.2 #&#
\begin{eqnarray}
\label{FLMZC}
&&E\bigl(Z_{\mathcal{T}(\tau)}(t) \mid Z_{\mathcal{S}(\tau)}, C=c\bigr)
\nonumber
\\[-8pt]
\\[-8pt]
\nonumber
&&\qquad =
\mu^{(c)}(t) + \int_{\mathcal{S}(\tau)}\beta_\tau^{(c)}(s,t)
\bigl\{Z_{\mathcal{S}(\tau)}(s)-\mu^{(c)}(s)\bigr\} \,ds
\end{eqnarray}
for all $t\in{\mathcal{T}(\tau)}$. Here, given a fixed value of
$\tau$, assume the
bivariate regression~function $\beta_\tau^{(c)}(s,t)$ is smooth and
square integrable, that is,\break $\int_{{\mathcal{T}(\tau)}} \int_{{\mathcal{S}(\tau)}}
\beta_\tau^{(c)}(s, t)\,ds \,dt < \infty$.
Under the smoothness assumption on the underlying random process, we
further assume that the bivariate regression function $\beta_\tau^{(c)}(s,t)$ is a smooth function of $\tau$ for all $s$ and $t$.
Using the eigenbasis expansion for the regression coefficient function
such that
$\beta_\tau^{(c)}(s,t)=\sum_{k=1}^\infty\sum_{j=1}^\infty
\beta_{\tau,kj}^{(c)} \varphi_{{\mathcal{S}(\tau)},j}^{(c)}(s) \varphi_{{\mathcal{T}(\tau)},k}^{(c)}(t)$,
model~(\ref{FLMZC}) can be expressed as
%
%
%e3.3 #&#
\begin{equation}\qquad
\label{FLMZC1} E\bigl(Z_{\mathcal{T}(\tau)}(t) \mid Z_{\mathcal{S}(\tau)}, C=c\bigr) =
\mu^{(c)}(t) + \sum_{j=1}^\infty\sum
_{k=1}^\infty\beta_{\tau
,kj}^{(c)}
\xi_{{\mathcal{S}(\tau)},j}^{(c)} \varphi_{{\mathcal
{T}(\tau)},k}^{(c)}(t),
\end{equation}
where $\xi_{{\mathcal{S}(\tau)},j}^{(c)}=\langle Z_{\mathcal
{S}(\tau)}-\mu^{(c)}, \varphi_{{\mathcal{S}(\tau)},j}^{(c)}
\rangle_{{\mathcal{S}(\tau)}}$ and
$\beta_{\tau,kj}^{(c)}={E(\xi_{{\mathcal{S}(\tau)},j}^{(c)} \xi_{{\mathcal{T}(\tau)}
,k}^{(c)})}/ \break {E\{(\xi_{{\mathcal{S}(\tau)},j}^{(c)})^2\}}$
are the regression parameters to be estimated.
Under the smoothness assumption on $\beta_\tau^{(c)}(s,t)$ along with
$\tau$, it follows that $\beta_{\tau,kj}^{(c)}$ is also smooth in
$\tau
$ for all $k$ and $j$.

%s3.2 #&#
\subsection{Functional linear prediction model for future traffic flow}\label{sec3,2}

Given $Z_{{\mathcal{S}(\tau)}}^*$, we aim to predict the values of
$Z_{{\mathcal{T}(\tau)}}^*$. Suppose
that the cluster structures $\mu^{(c)}$ and $\{\varphi_{{\mathcal
{S}(\tau)},j}^{(c)}\}$\vadjust{\goodbreak}
and the regression coefficients $\beta_{\tau,kj}^{(c)}$ are given. In
practice, these estimates can be obtained from the functional
clustering and the functional regression analysis using the historical
or training data as described in Section \ref
{clustering&classification}. Then, the functional prediction model
below is used to predict the
unobserved trajectory conditional on a specific cluster:
%
%
%e3.4 #&#
\begin{equation}
\quad E\bigl(Z^*_{\mathcal{T}(\tau)}(t) \mid Z^*_{\mathcal{S}(\tau)}, C=c\bigr) =
\mu^{(c)}(t) + \sum_{j=1}^\infty\sum
_{k=1}^\infty\beta_{\tau
,kj}^{(c)}
\xi_{{\mathcal{S}(\tau)},j}^{*(c)} \varphi_{{\mathcal
{T}(\tau)},k}^{(c)}(t)
\label{FLMZ*C1}
\end{equation}
for all $t\in{\mathcal{T}(\tau)}$,
where $\xi_{{\mathcal{S}(\tau)},j}^{*(c)}=\langle Z_{\mathcal
{S}(\tau)}^{*}-\mu^{(c)}, \varphi_{{\mathcal{S}(\tau)}
,j}^{(c)} \rangle_{{\mathcal{S}(\tau)}}$ and will be obtained by
numerical approximation.

Finally, given a partially observed trajectory $Z^*_{{\mathcal{S}(\tau
)}}$, the
unobserved trajectory $Z^*_{{\mathcal{T}(\tau)}}$ can be predicted by
the functional
mixture prediction model~(\ref{mixturepred}) using the results of the
functional prediction model~(\ref{FLMZ*C1}) in conjunction with the
multiclass logit model~(\ref{mclogitK*}).
However, the components in these models remain to be estimated.
The estimation procedure for the functional linear model is briefly
summarized below.

%s3.3 #&#
\subsection{Estimation for functional mixture prediction models}\label{sec3,3}
We note that the estimation of $\beta_{\tau,kj}$ in~(\ref{FLMZC1})
and~(\ref{FLMZ*C1}) can further be simplified using a simple linear
regression approach [M\"uller, Chiou and Leng (\citeyear{MulChiLen08})],
such that
\[
E\bigl(\xi_{{\mathcal{T}(\tau)},k}^{(c)} \mid\xi_{{\mathcal{S}(\tau
)},j}^{(c)}
\bigr)= \beta_{\tau,kj}^{(c)} \xi_{{\mathcal{S}(\tau)},j}^{(c)}
\]
for all pairs of $(k,j)$.
Therefore, functional linear regression can be decomposed into a series
of simple linear regressions of functional principal component scores
of the response processes in relation to those of the predictor processes.

For our predictions, given the cluster membership information and the
subspace structure of each cluster, we estimate $\beta_\tau^{(c)}(s,t)$
in the functional linear regression model~(\ref{FLMZC}) based on the
training data.
Given the estimated principal component functions $\hat\varphi_{{\mathcal{S}(\tau)}
,j}^{(c)}(t)$ and $\hat\varphi_{{\mathcal{T}(\tau)},k}^{(c)}(t)$
and the principal
component scores $\hat\xi_{{\mathcal{S}(\tau)},j}^{(c)}$ and $\hat
\xi_{{\mathcal{T}(\tau)}
,k}^{(c)}$, the
estimate of $\beta_{\tau,kj}^{(c)}$ can be obtained by
%
%
%e3.5 #&#
\begin{equation}
\label{tildebetahat} \qquad\tilde\beta_{\tau,kj}^{(c)} = \bigl
\{(n_c-1)\lambda_{{\mathcal
{S}(\tau)}
,j}^{(c)} \bigr\}^{-1}
\sum_{i=1}^{n_c} \bigl(\hat
\xi_{{\mathcal{S}(\tau
)},i,j}^{(c)}-\overline{\xi }_{{\mathcal{S}(\tau)}
,j}^{(c)}
\bigr) \bigl(\hat\xi_{{\mathcal{T}(\tau
)},i,k}^{(c)}-\overline{\xi}_{{\mathcal{T}(\tau)}
,k}^{(c)}
\bigr),
\end{equation}
where $\overline{\xi}_{{\mathcal{S}(\tau)},j}^{(c)}$ and $\overline
{\xi}_{{\mathcal{T}(\tau)},k}^{(c)}$
are sample averages of $\hat\xi_{{\mathcal{S}(\tau)},i,j}^{(c)}$
and $\hat\xi_{{\mathcal{T}(\tau)}
,i,k}^{(c)}$, respectively.

To take advantage of smoothness in prediction as the value $\tau$
progresses, we further smooth the estimates $\{\tilde\beta_{\tau
,kj}^{(c)}, \tau= \tau_1, \ldots, \tau_Q\}$ over $\tau$ to obtain the
smooth estimates $\hat\beta_{\tau,kj}^{(c)}$, where $Q$ is the number
of time points at which predicting the future trajectory is of
interest. Here,\vadjust{\goodbreak} we use the local linear smoothing method with
cross-validated bandwidth [see, e.g., \citet{FanGij96}].
Accordingly, using $\hat\beta_{\tau,kj}^{(c)}$ and the estimates
$\hat
\mu^{(c)}(t)$ and $\hat\varphi_{{\mathcal{T}(\tau)},k}^{(c)}(t)$,
we obtain the
predicted trajectory conditional on cluster $c$ by
%
%
%e3.6 #&#
\begin{eqnarray}
\label{FLMZ*Chat} \hat{Z}^{*(c)}_{\mathcal{T}(\tau)}(t)&=&\hat{E}
\bigl(Z^*_{\mathcal
{T}(\tau)}(t) \mid Y^*_{\mathcal{S}(\tau)}, C=c\bigr)
\nonumber
\\[-8pt]
\\[-8pt]
\nonumber
&=& \hat
\mu^{(c)}(t) + \sum_{j=1}^{M_c} \sum
_{k=1}^{M_c} \hat\beta_{\tau
,kj}^{(c)}
\hat\xi_{{\mathcal{S}(\tau)},j}^{*(c)} \hat\varphi_{{\mathcal{T}(\tau)},k}^{(c)}(t)
\end{eqnarray}
for all $t\in{\mathcal{T}(\tau)}$.
Here, $M_c$ is determined by~(\ref{Mc}).
Finally, combining the results of~(\ref{FLMZ*Chat}) with~(\ref
{mclogitK*hat}), we obtain the predicted
unobserved traffic flow trajectory
%
%
%e3.7 #&#
\begin{equation}
\label{mixturepredhat} \hat{Z}^{*}_{\mathcal{T}(\tau)}(t)=\hat E
\bigl(Z_{{\mathcal{T}(\tau
)}}^*(t) \mid Y_{{\mathcal{S}(\tau)}}^*\bigr) = \sum
_{c=1}^K \hat{Z}^{*(c)}_{\mathcal{T}(\tau)}(t)
\hat P\bigl(C=c \mid Y_{{\mathcal{S}(\tau)}}^*\bigr) .
\end{equation}

%s3.4 #&#
\subsection{Implementation algorithm of functional mixture predictions}\label{sec3.4}

Suppose there is a newly observed trajectory $\{(t^*_{j},Y^*(t^*_{j}));
t^*_{j} < \tau\}$, denoted by $Y^*_{{\mathcal{S}(\tau)}}$ for short.
The algorithm for functional mixture prediction that combines the
functional classification procedure with the functional prediction
model is summarized as follows.
\begin{enumerate}[]
\item[Step 1.] \textit{Identification of cluster subspaces.} Perform the
subspace-projected functional clustering procedure according to
criterion~(\ref{ccondprob}) to identify cluster subspaces, $\{ \hat
\mu^{(c)}, \{\hat\varphi_k^{(c)}\}_{k=1, \ldots, M_c} \}$, $c=1, \ldots
, K$,
based on the training data set as discussed in Sections~\ref
{clustering} and~\ref{estimationclustering}.

\item[Step 2.] \textit{Model fitting based on the historical or
training data.}
\begin{enumerate}[(ii)]
\item[(i)] \textit{Obtain the multiclass logit model for cluster
membership distributions.}
Obtain from Step 1 the regression coefficient estimates $\hat{\bgamma
}_{c}$ in~(\ref{mclogitKhat}).
\item[(ii)] \textit{Fit the functional linear regression model.} Fit
the cluster-specific functional linear regression models and obtain the
regression coefficient estimates $\hat\beta_{\tau,kj}^{(c)}$ as a
smoothed version of~(\ref{tildebetahat}).
\end{enumerate}

\item[Step 3.] \textit{Prediction of the future traffic flow trajectory
for a new and partially observed $Y^*_{{\mathcal{S}(\tau)}}$
conditional on clusters.}
\begin{enumerate}[(ii)]
\item[(i)] \textit{Predict the posterior membership probability of
$Y^*_{{\mathcal{S}(\tau)}}$ associated with each cluster.} Calculate
the relative $L^2$
distances $d_{\mathcal{S}(\tau)}^{*(c)}$ for the given
$Y^*_{{\mathcal{S}(\tau)}}$ in~(\ref{dSY*})
and obtain the posterior probability $\hat{\operatorname{Pr}}(C=c\mid Y_{\mathcal
{S}(\tau)}^*)$
in~(\ref{mclogitK*hat}).
\item[(ii)] \textit{Predict the
unobserved trajectory $Y_{\mathcal{T}(\tau)}^*$ conditioning on each
of the clusters.}
Obtain the cluster-specific functional prediction model fit $\hat
{E}(Y_{\mathcal{T}(\tau)}^*\mid Y_{\mathcal{S}(\tau)}^*,C=c)$
in~(\ref{FLMZ*Chat}).
\end{enumerate}

\item[Step 4.] \textit{Prediction of traffic flow trajectory by the
functional mixture prediction model.} Calculate the predicted
trajectory $\hat{E}(Y_{\mathcal{T}(\tau)}^*\mid Y_{\mathcal{S}(\tau
)}^*)$ in~(\ref{mixturepredhat})
using the results of $\hat{\operatorname{Pr}}(C=c\mid Y_{\mathcal{S}(\tau)}^*)$ and
$\hat{E}(Y_{\mathcal{T}(\tau)}^*\mid
Y_{\mathcal{S}(\tau)}^*,C=c)$ and obtain the bootstrap prediction intervals.
\end{enumerate}
Details in constructing the bootstrap prediction intervals in Step 4
are provided in Supplementary Material B [\citet{chi12}].

%s4 #&#
\section{Analysis of traffic flow patterns}\label{sec4}

The sample data set of daily traffic flow trajectories from
Section~\ref
{subsectraffic} is divided into a training data set (70 days) and a
test data set (14 days) to examine the predictive performance of our
model. Clusters of the traffic flow patterns from the training data are
identified based on subspace projection using the proposed
subspace-projected functional clustering method according to
criterion~(\ref{ccondprob}).
The implementation of the \textit{functional forward testing procedure}
of \citet{LiChi11} leads to the choice of 3 clusters.
Table~\ref{tabclusternumber} summarizes the empirical probabilities of
rejecting the null hypotheses for $K=2, 3, 4$, based on 200 bootstrap
samples. The $p$-values with reference to the predetermined level of
significance 0.05, adjusted for multiple comparisons, indicate that
when $K=2$ and 3, the clusters are all significantly distinct, while
Clusters 1 and 4 when $K=4$ are not significantly different in terms of
the mean functions and the eigenspaces.
%
%
%t1 #&#
\begin{table}[b]
\caption{Empirical probabilities of rejecting the null hypotheses
$H_{01}$ and $H_{02}$, respectively, based on 200 bootstrap samples}
\label{tabclusternumber}
\begin{tabular*}{\textwidth}{@{\extracolsep{\fill}}lccc@{}}
\hline
\textbf{Number of clusters} & \textbf{Clusters} & \multicolumn{1}{c}{$\bolds{H_{01}\dvtx \mu^{(c)}=\mu^{(d)}}$} &
\multicolumn{1}{c@{}}{$\bolds{H_{02}\dvtx S^{(c)}=S^{(d)}}$} \\
\hline
$2$ & 1 vs. 2 & $0.000$ & $0.010$\\[3pt]
{$3$} & 1 vs. 2 & $0.000$ & $0.005$\\
& 1 vs. 3 & $0.000$ & $0.005$\\
& 2 vs. 3 & $0.005$ & $0.015$\\[3pt]
{$4$} & 1 vs. 2 & $0.000$ & $0.055$\\
& 1 vs. 3 & $0.000$ & $0.030$\\
& 1 vs. 4 & $0.155$ & $0.025$\\
& 2 vs. 3 & $0.000$ & $0.160$\\
& 2 vs. 4 & $0.000$ & $0.035$\\
& 3 vs. 4 & $0.000$ & $0.000$\\
\hline
\end{tabular*}
\end{table}

%
%f3 #&#
\begin{figure}

\includegraphics{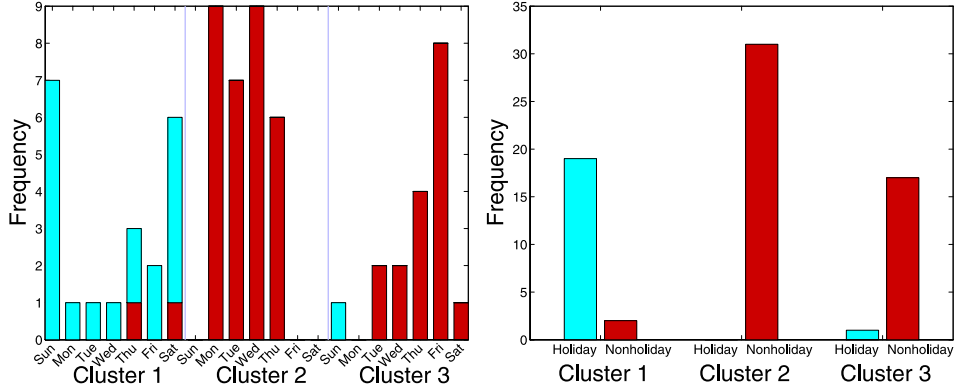}

\caption{Frequency plots of cluster labels by days of the week
(left panel) and by non- and holidays (right panel) for Clusters 1, 2
and 3 (left, middle and right groups).}
\label{figFlow-day-holi}
\end{figure}

%
%f4 #&#
\begin{figure}[b]

\includegraphics{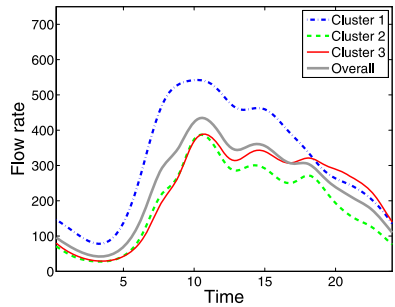}

\caption{Overall and cluster-specific mean functions of the
training data of daily traffic flow rates.}
\label{figOverall-mean-functions}
\end{figure}

%
%f5 #&#
\begin{figure}

\includegraphics{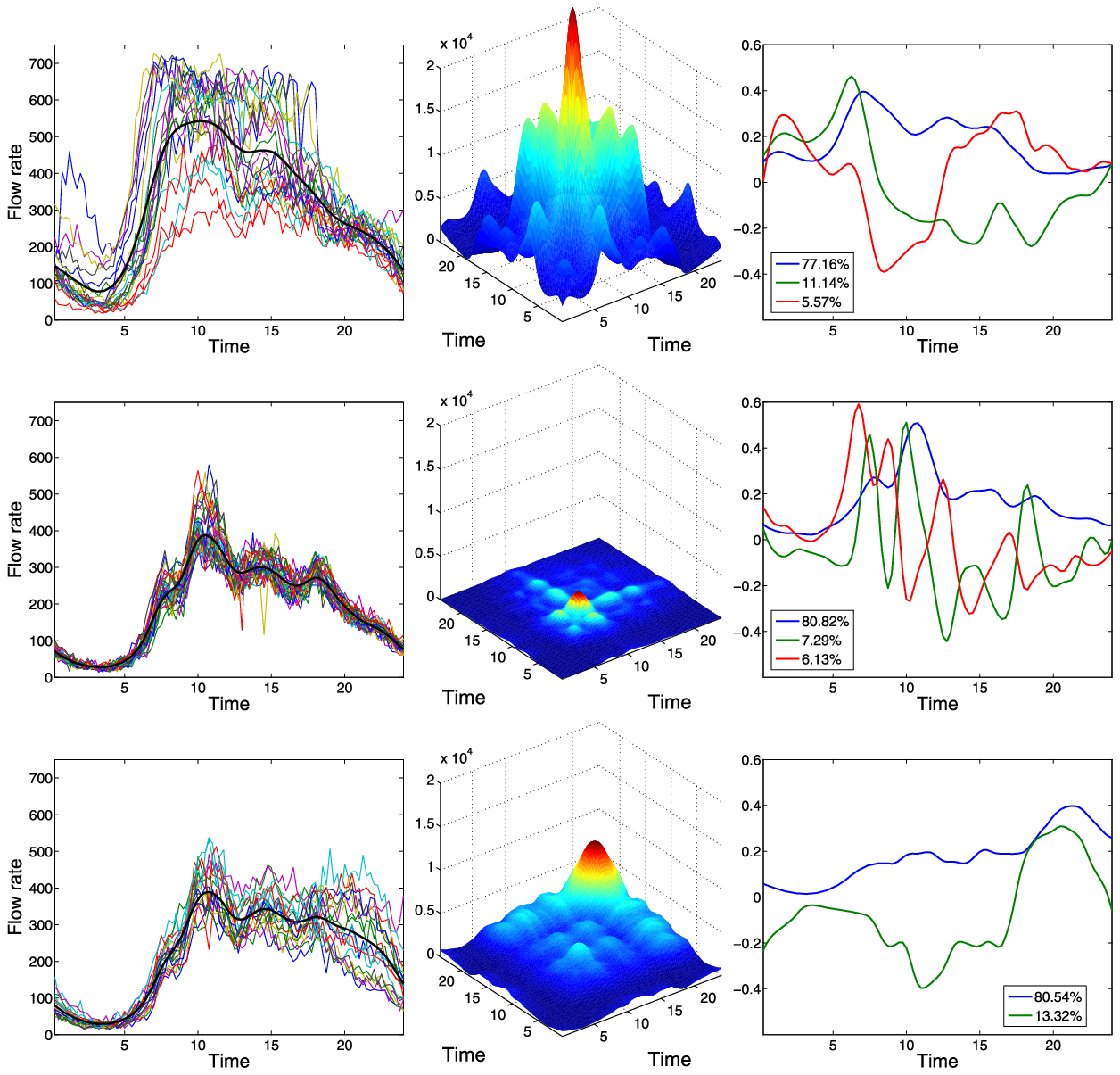}

\caption{Estimated mean functions (left column) superimposed on
the observed trajectories, covariance functions (middle column) and the
corresponding eigenfunctions (right column) of Clusters 1--3 (from top
to bottom) based on the training data of daily traffic flow trajectories.}
\label{figmean-cov-eig-c1-c3}
\end{figure}

The cluster memberships displayed in Figure~\ref{figFlow-day-holi}
show that Cluster 1 contains mostly weekends (left panel), with 90\%
being holidays including weekends (right panel). Cluster 2 completely
comprises weekdays including Mondays through Thursdays. Cluster 3
comprises mostly weekdays, especially Fridays (left panel).
The mean functions of the three clusters and the overall trajectories
are displayed in Figure~\ref{figOverall-mean-functions}. While Cluster
1 has a higher mean traffic flow rate\vadjust{\goodbreak} than the other two clusters,
Clusters~2 and 3 have relatively close mean flow rates in terms of
shape and magnitude until 11:00, and they diverge thereafter with a
higher mean flow rate in Cluster 3. The observed trajectories along
with the corresponding covariance functions and leading eigenfunctions
are shown in Figure~\ref{figmean-cov-eig-c1-c3}. The variability of
Cluster~1 is higher than the other two clusters, while Cluster~2 has
the lowest variability. The peak flow rate in Cluster 1 lasts from
07:00 to 17:00 and the three leading principal component functions
explain 77.16\%, 11.14\% and 5.57\% of total variability. The
trajectories in Cluster 2 have a relatively uniform pattern with the
major peak flow rate at around 11:00. Cluster~3 indicates a high
variability of flow rates occurring after 18:00.
The mean integrated prediction errors are defined as
${n_c}^{-1}\sum_{i=1}^{n_c}{T}^{-1}\int_{0}^{T} (\hat
{Z}_{i}^{(c)}(t)-Y_{i}(t) )^2\,dt$,
where $\hat{Z}_{i}^{(c)}(t)=\hat\mu^{(c)}(t)+\sum_{j=1}^{M_c} \hat
\xi_{ij}^{(c)} \hat\varphi_{j}^{(c)}(t)$, $Y_i(t)$ is the observed
trajectory and $n_c$ is the number of trajectories in Cluster $c$.
These are 327.3, 78.8 and 122.4 for Clusters 1--3, respectively.
Prediction using the overall trajectories without clustering, in
contrast, returned an error of 300.5, indicating that there is a huge
reduction in prediction errors when heterogeneity of cluster patterns
are taken into account.

%
%f6 #&#
\begin{figure}

\includegraphics{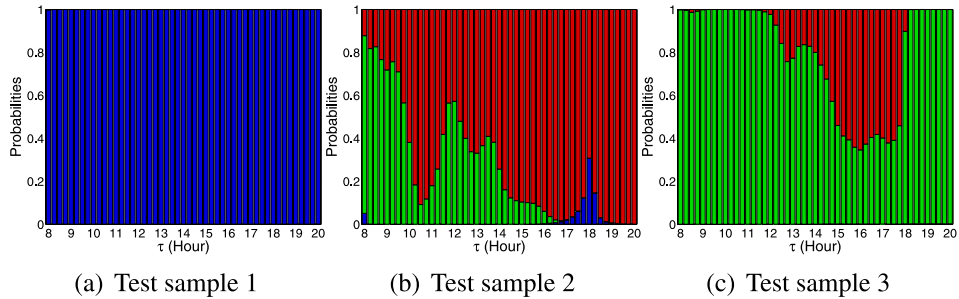}

\caption{The predicted cluster membership distribution for
Clusters 1--3 (indicated in blue, green and red) as a function of the
``current time'' $\tau$ (per 15 min) for samples from the test data based
on the trajectories observed up to $\tau$.}
\label{figprob3cidv}
\end{figure}

The model fits of the multiclass logistic regression listed in
Supplementary Material C [\citet{chi12}]
are used in predicting the
unobserved trajectory for an up-to-date and partially observed trajectory.
Given a newly observed trajectory from the test data up to the current
time $\tau$, to consider different flow patterns, we predict the
posterior probabilities for each of the associated clusters by
functional classification based on the multiclass logistic regression
model in~(\ref{mclogitK*hat}), using the fitted regression
coefficients in~(\ref{mclogitKhat})
with the relative $L^2$ distances~(\ref{dSY*}) as the covariate. The
posterior probabilities for some test samples are illustrated in
Figure~\ref{figprob3cidv}
with the values of $\tau$ progressing from 08:00 to 20:00 by 15-min intervals.
In Figure~\ref{figprob3cidv}(a), the predicted membership
probabilities of \textit{Test sample 1}
degenerate to one for all values of $\tau$. The associated predicted
trajectories are shown in Figure~\ref{figFlowctx0cty0-sub3}(a)--(c).
The time-varying prediction intervals, which are wider closer to $\tau$
and taper off toward the end, depend on Cluster 1's variability pattern
as illustrated in Figure~\ref{figmean-cov-eig-c1-c3} (top panels). In
contrast, Figure~\ref{figprob3cidv}(b) for \textit{Test sample 2}
indicates a more complex situation where the predicted membership
distributions change with~$\tau$, with the associated predicted
trajectories shown in Figure~\ref{figFlowctx0cty0-sub3}(d)--(f). In this case, using the early
trajectory information up to $\tau=8\mbox{:}00$ may lead to misclassification,
which makes it difficult to predict its future trajectory accurately.
This issue is resolved as $\tau$ moves onward.
The wider prediction bands with $\tau$ at 12:00 and 16:00, in
comparison with that at 8:00, reflect the fact that the variability of
Cluster 3's traffic flow trajectories is larger in the afternoon
as illustrated in Figure~\ref{figmean-cov-eig-c1-c3} (bottom panels).
Figure~\ref{figprob3cidv}(c) indicates that the posterior cluster
membership probabilities of \textit{Test sample 3} alternate between
Clusters 2 and 3, owing to certain similarities in these two cluster
patterns, and
the predicted cluster membership remains with Cluster 3
after around 18:00.
Given that the actual cluster memberships are unknown, the accuracy of
functional classification for the up-to-date and partially observed
trajectories in the test data will be investigated via a simulation
study in Section \ref{sec6}.

%s5 #&#
\section{Traffic flow prediction}\label{sec5}

In predicting the
unobserved traffic flow trajectories, we also investigate the effects
on the prediction performance of the interval length prior to time
$\tau
$ and the future interval length after it. To this end, we define
${\mathcal{S}(\tau; \omega)}
= [\max(0,\tau-\omega),\tau]$, where $\omega$ is the length of the
known interval to be used in prediction calculations and ${\mathcal
{T}(\tau; \kappa)}= [\tau,
\min(\tau+\kappa,T)]$,
where $\kappa$ is the length of the unknown interval to be predicted
from time $\tau$ onward.
In the test data, given a sample $Y_i^*$ observed up to time $\tau$,
denoted by $Y_{i,{\mathcal{S}(\tau)}}^*$, we define the mean
integrated prediction error
($\operatorname{MIPE}$) as the the performance measure of predicting $Y_{i,{\mathcal
{T}(\tau)}}^*$.
This is expressed as
$\operatorname{MIPE}(\tau,\omega,\kappa) = m_p^{-1} \sum_{i=1}^{m_p} \kappa^{-1}
\int_{0}^{\kappa}  \{ \widehat{Z}_{i,{\mathcal{T}(\tau)}}^*(t) -
Y_{i,{\mathcal{T}(\tau)}}^*(t)
\}^2 \,dt$,
where $Y_{i,{\mathcal{T}(\tau)}}^*(t)=Y_{i}^*(\tau+t)$, $\widehat
{Z}_{i,{\mathcal{T}(\tau)}}^*(t)$ is
obtained by~(\ref{mixturepredhat}) and $m_p$ is the number of
trajectories in the test data.
For ease of comparisons across different values of $\tau$, let $\tau_s=\max(0,\tau-\omega)$ and $\tau_e=\min(\tau+\kappa,T)$, for
$\omega>
0$ and $\kappa> 0$.
We define the total mean integrated prediction error ($\operatorname{TMIPE}$)
for the overall prediction performance by
%
%
%e5.1 #&#
\begin{equation}
\label{TMIPE} \operatorname{TMIPE}(\omega,\kappa) = \int_{\tau_s}^{\tau_e}
\operatorname{MIPE}(\tau,\omega ,\kappa) \,d\tau,
\end{equation}
where $\tau_s$ and $\tau_e$ are the smallest and the largest values,
respectively, selected with respect to the times, $\tau$, on the
domain $[0,T]$.
In this study, $T=24$ (hours) and we set $\tau_s=8$ and $\tau_e=20$.
For notational convenience, we let $\kappa^*=24-\tau$ and $\omega^*=\tau
$, $\tau_s\le\tau\le\tau_e$, such that $\kappa^*$ denotes the interval
length from the current time to the end of the day and $\omega^*$
denotes the maximal length of the past trajectory information available
for prediction.

%s5.1 #&#
\subsection{Results and comparisons of traffic flow prediction}\label{sec5,1}

In this study we investigate the prediction performance by comparing
the following methods:
\begin{itemize}
\item$\mathrm{FP}$: Functional prediction based on functional linear
regression using the same setting described in Section~\ref{FLR} but
without considering clusters of different traffic flow patterns;
\item$\mathrm{FMP}_{H}$: Functional prediction using the proposed
functional mixture prediction model except that the posterior
membership probabilities~(\ref{mclogitK*}) degenerate to zero or one
such that $\sum_{c=1}^K P(C=c \mid Z_{\mathcal{S}(\tau)}^*)=1$
(where the subscript $H$
reflects the so-called \textit{hard} classification);
\item$\mathrm{FMP}_{S}$: Functional prediction using the proposed
functional mixture prediction model (where the subscript $S$ reflects
the so-called \textit{soft} classification or probabilistic classification).
\end{itemize}

%
%t2 #&#
\begin{table}
\caption{Performance comparisons for $\mathrm{FP}$, $\mathrm{FMP}_H$
and $\mathrm{FMP}_S$ based on
$\operatorname{TMIPE}$ ($\times10^3$) under~various values of $\kappa$ and $\omega$}
\label{tabaucflm3VD}
\begin{tabular*}{\textwidth}{@{\extracolsep{\fill}}lcd{2.2}d{2.2}d{2.2}d{2.2}d{2.2}d{2.2}d{2.2}@{}}
\hline
&&\multicolumn{7}{c@{}}{$\bolds{\omega}$}\\[-6pt]
&&\multicolumn{7}{c@{}}{\hrulefill}\\
& $\bolds{\kappa}$ & \multicolumn{1}{c}{\textbf{1}} & \multicolumn{1}{c}{\textbf{2}} & \multicolumn{1}{c}{\textbf{3}} &
\multicolumn{1}{c}{\textbf{4}} & \multicolumn{1}{c}{\textbf{5}} & \multicolumn{1}{c}{\textbf{6}} & \multicolumn{1}{c@{}}{$\bolds{\omega^{*}}$}\\
\hline
{$\mathrm{FP}$}
& 1 & 4.12 & 4.92& 4.84& 4.92& 5.30& 5.68& 5.82\\
& 4 & 7.42 & 7.71& 7.69& 8.10& 8.60& 9.06& 8.95\\
& 8 & 9.92 &10.26&10.35&10.79&11.27&11.71&11.65\\
& $\kappa^{*}$ & 12.34&12.79&12.94&13.41&13.91&14.36&14.30\\[3pt]
{$\mathrm{FMP}_{H}$}
& 1 & 3.48&3.22&3.20&3.33&3.53&3.52&3.14\\
(3 clusters)& 4 & 5.00&4.62&4.73&4.91&5.20&5.26&4.79\\
& 8 & 8.88&8.44&8.48&8.68&8.97&9.07&8.81\\
& $\kappa^{*}$ & 12.24&11.82&11.87&12.06&12.36&12.47&12.33\\[3pt]
{$\mathrm{FMP}_{S}$ }
& 1 & 3.31&3.26&3.07&2.93&2.87&2.80&2.81\\
(3 clusters) & 4 & 4.37&4.14&4.05&3.98&3.80&3.86&4.18\\
& 8 & 5.80&5.64&5.54&5.49&5.41&5.61&6.68\\
& $\kappa^{*}$ & 7.97&7.94&7.84&7.86&7.71&7.89&9.95\\
\hline
\end{tabular*}
\end{table}

%
%f7 #&#
\begin{figure}[b]

\includegraphics{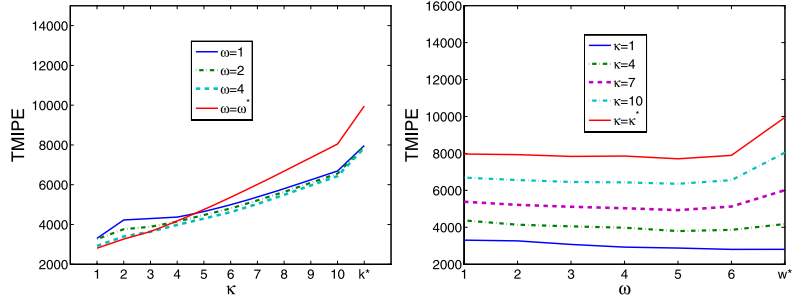}

\caption{Performance comparisons for $\mathrm{FMP}_{S}$, based on
$\operatorname{TMIPE}$ (\protect\ref{TMIPE}),
displayed as a function of~$\kappa$ (left) with $\omega$ fixed at 1, 2,
4 and $\omega^*$ and as a function of
$\omega$ (right) with $\kappa$ fixed at 1, 4, 7, 10 and $\kappa^*$.}
\label{figaucmlogitKCFC3ofFlow}
\end{figure}

\begin{figure}

\includegraphics{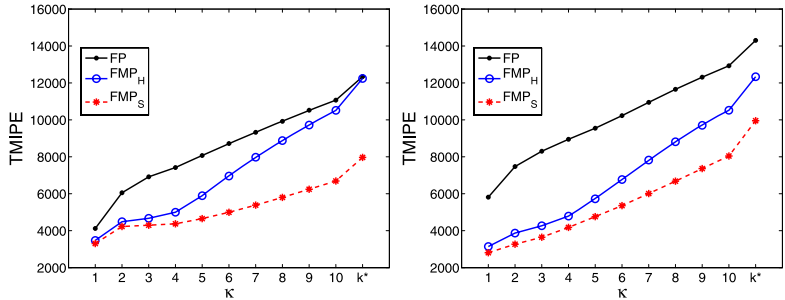}

\caption{Performance comparisons for $\mathrm{FP}$, $\mathrm{FMP}_{H}$
and $\mathrm{FMP}_{S}$,
based on $\operatorname{TMIPE}$ (\protect\ref{TMIPE}), displayed as a function of
$\kappa$,
with $\omega=1$ (left) and $\omega=\omega^*$ (right).}
\label{figAUCFLM}
\end{figure}

To examine prediction performance under various situations, we consider
a wide range of values $\tau$ along with various values of $\omega$ and
$\kappa$ as defined in ${\mathcal{S}(\tau; \omega)}$ and ${\mathcal
{T}(\tau; \kappa)}$.
Table~\ref{tabaucflm3VD} indicates that the proposed $\mathrm
{FMP}_{S}$ is robust, generally outperforming the other two ($\mathrm
{FP}$ and $\mathrm{FMP}_{H}$) under various values of $\omega$ and
$\kappa$.
Figure~\ref{figaucmlogitKCFC3ofFlow} (left panel) indicates that
small values of $\omega$ $(1,2,4)$ give a similar performance and it is
not surprising that the performance for larger values of $\kappa$ is
worse. For fixed values of $\kappa$ (right panel), TMIPE as a function
of $\omega$ generally shows a positive slope as it moves away from the
origin, with a minimum at $\omega=5$ (for $\kappa=4,8,\omega^*$) or
$\omega=6$ (for $\kappa=1$) from Table~\ref{tabaucflm3VD}. The trend
is relatively flat, but
becomes steeper when
$\omega=\omega^*$.
This discrepancy is more pronounced with increasing $\kappa$. A
possible explanation is that the flow trajectory patterns in Clusters 2
and 3 are close in shape and magnitude until noon and diverge
thereafter and, thus, using larger $\omega$ with more past information
may not significantly improve the overall prediction accuracy.
In the literature, Sent\"urk and M\"uller (\citeyear{Se10}) considered
the length
of past data to be used for prediction and suggested the optimal length
using a data-adaptive criterion that minimizes the absolute prediction error.
Our empirical results also suggest the use of a data-adaptive criterion
to choose the length of past data.
Additionally, comparisons for the prediction performance between the
methods with fixed values of $\omega$ as illustrated in Figure~\ref
{figAUCFLM} (with $\omega=1$ on the left and $\omega=\omega^*$ on the
right) reinforce the conclusion that $\mathrm{FMP}_{S}$ outperforms
$\mathrm
{FP}$ and $\mathrm{FMP}_{H}$.
In addition to the 3-cluster prediction performance illustrated above,
results of the 2- and 4-cluster prediction performances are illustrated
in Supplementary Material~C [\citet{chi12}] for comparisons. These results also support
our choice of the 3-cluster model, which outperforms the 2- and
4-cluster models.

%
%f8 #&#

%s5.2 #&#
\subsection{Comparisons with other methods}\label{sec5,2}

We compare the functional mixture prediction approach with an existing
method that could also be fitted into our functional mixture prediction
framework. One possible approach is to treat the unobserved future
trajectory for a partially observed trajectory as missing in the entire
trajectory. That is, we may replace~(\ref{FLMZ*Chat}) in Step~3(ii)
with the model
%
%
%e5.2 #&#
\begin{equation}
\label{FPCPZ*Chat} \hat{Z}^{*(c)}_{\mathcal{T}(\tau)}(t)=\hat{E}
\bigl(Z^*_{\mathcal
{T}(\tau)}(t) \mid Y^*_{\mathcal{S}(\tau)}, C=c\bigr) = \hat
\mu^{*(c)}(t) + \sum_{j=1}^{M_c} \hat
\xi_{j}^{*(c)} \hat \varphi_{j}^{*(c)}(t)
\end{equation}
for all $t\in{\mathcal{T}(\tau)}$, where the estimated mean function
$\hat\mu^{*(c)}(t)$
and the estimated eigenfunctions of the covariance kernel $\hat\varphi_{j}^{*(c)}(t)$ of cluster $c$ are obtained by the training data set in
the clustering step with the corresponding domain ${\mathcal{T}(\tau
)}$. The key step is
to estimate the functional principal component scores, $\hat\xi_{j}^{*(c)}$, which cannot be obtained easily since the trajectory is
only partially observed.
An existing method that can deal with this situation makes use of the
expectation of the posterior distribution in Proposition 1 of
\citet{ZhoSerGeb11}, assuming that the prior distribution of the
scores is Gaussian, for an application to degradation modeling. This
formula coincides with the conditional expectation approach in equation
(4) of Yao, M\"uller and Wang (\citeyear{YaoMulWan05N1}) under
Gaussian assumptions,
although they are different in terms of statistical inference. We term
this method the Functional Principal\vadjust{\goodbreak} Component Prediction (FPCP)
approach. We apply FPCP to the proposed functional mixture prediction
algorithm including the cases with and without
clustering/classification considerations for comparisons, including
$\mathrm{FPCP}$, $\mathrm{FPCP}_{H}$ and $\mathrm{FPCP}_{S}$ that are
parallel to $\mathrm{FP}$, $\mathrm{FMP}_{H}$ and $\mathrm{FMP}_{S}$. The
results shown in Table~\ref{tabaucfpc}, in comparison with the
results in Table~\ref{tabaucflm3VD}, indicate that the functional
mixture prediction approach in conjunction with functional linear
regression outperforms the FPCP approach. Additional results for the 2-
and the 4-cluster models are also provided in Supplementary Material C [\citet{chi12}]
for comparison.

%
%t3 #&#
\begin{table}
\caption{Performance comparisons for $\mathrm{FPCP}$, $\mathrm{FPCP}_H$
and $\mathrm{FPCP}_S$ based on $\operatorname{TMIPE}$~($\times10^3$) under various
values of $\kappa$ and $\omega$}
\label{tabaucfpc}
\begin{tabular*}{\textwidth}{@{\extracolsep{\fill}}lcd{2.2}d{2.2}d{2.2}d{2.2}d{2.2}d{2.2}d{2.2}@{}}
\hline
&&\multicolumn{7}{c@{}}{$\bolds{\omega}$}\\[-6pt]
&&\multicolumn{7}{c@{}}{\hrulefill}\\
& $\bolds{\kappa}$ & \multicolumn{1}{c}{\textbf{1}} & \multicolumn{1}{c}{\textbf{2}} & \multicolumn{1}{c}{\textbf{3}} &
\multicolumn{1}{c}{\textbf{4}} & \multicolumn{1}{c}{\textbf{5}} & \multicolumn{1}{c}{\textbf{6}} & \multicolumn{1}{c@{}}{$\bolds{\omega^{*}}$}\\
\hline
{$\mathrm{FPCP}$ }
& 1 & 14.30&13.94&13.63&12.62&10.49&8.48&7.69\\
& 4 & 17.43&16.70&15.37&13.57&11.90&11.11&10.77\\
& 8 & 17.97&16.98&15.96&14.93&14.10&13.73&13.30\\
& $\kappa^{*}$ & 19.27&18.59&17.88&17.13&16.50&16.19&15.66\\[3pt]
{$\mathrm{FPCP}_H$}
& 1 & 7.83&8.62&8.90&8.90&10.57&10.67&2.85\\
(3 clusters)& 4 &10.72&11.35&11.99&11.63&11.21&9.75&4.80\\
& 8 &12.57&12.62&12.28&11.94&12.05&11.65&9.19\\
& $\kappa^{*}$ & 14.35&14.36&14.24&14.13&14.14&13.94&12.48\\[3pt]
{$\mathrm{FPCP}_S$}
& 1 & 6.26&7.29&7.54&7.73&9.41&9.44&3.07\\
(3 clusters)& 4 & 8.73&9.84&10.79&10.86&10.61&9.86&4.87\\
& 8 &10.13&11.16&11.66&11.65&11.60&11.28&8.80\\
& $\kappa^{*}$ &12.12&12.95&13.23&13.12&13.14&13.14&12.37\\
\hline
\end{tabular*}
\end{table}

%s6 #&#
\section{Simulation}\label{sec6}

We implement a Monte Carlo simulation to evaluate the performance of
the functional clustering and classification procedures as well as the
functional prediction accuracy. We simulate the scenario of the real
traffic flow trajectories analyzed in the previous sections. We
generate a training data set and a test data set for each simulation
run using the estimated results of the 3-cluster traffic flow
trajectories as the true models with a total of 100 simulation
replicates. The numbers of curves $n_c$ are 21, 31 and 18 for Clusters
1--3 in each training data set and are 3, 8 and 3 in each test data as
in the previous analysis. The synthetic curves of cluster $c$ are
generated by the model
$y_{i}^{(c)}(t_{j})=\mu^{(c)}(t_{j})+\sum_{j=1}^{\widetilde{M}_c}
\xi_{ij}^{(c)}\phi_{j}^{(c)}(t_{j}) +\varepsilon_{ij}^{(c)}$,
for $i=1, \ldots, n_c$, where $\xi_{ij}^{(c)}$ are normal random
variates with a mean of zero and variance $\lambda^{(c)}$ and the
random measurement errors $\varepsilon_{ij}^{(c)}$ are independent and
follow a normal distribution with a mean of zero and variance $\sigma^{2}_{(c)}$. The recording times $t_j=j/4$ for $j=1,\ldots,96$ mimic
the 15-min recording time interval. The quantities $\mu^{(c)}$, $\phi_j^{(c)}$, $\lambda_j^{(c)}$ and $\sigma^{2}_{(c)}$
use the model estimates of
our real traffic flow data analysis. The numbers of components
$\widetilde{M}_{c}$ are determined by the numbers of the estimated
$\lambda_j^{(c)}$ that are strictly positive. Further details of the
simulated models regarding the underlying functions $\mu^{(c)}$ and
$\phi_j^{(c)}$, along with a sample of synthetic trajectories, are
displayed in Supplementary Material D [\citet{chi12}]. The clustering results of this
simulated sample, including the estimated mean function and the
eigenfunctions along with the covariance functions, are also illustrated.

%
%t4 #&#
\begin{table}[b]
\caption{Average $\operatorname{TMIPE}$ ($\times10^3$) (with s.e. in parentheses)
for $\mathrm{FP}$, $\mathrm{FMP}_H$, $\mathrm{FMP}_S$ and $\mathrm{FMP}_S^*$
under various values of $\kappa$ and $\omega$ based on 100 simulation
replicates}
\label{tabaucflm3sim}
\begin{tabular*}{\textwidth}{@{\extracolsep{\fill}}lcccccc@{}}
\hline
&&\multicolumn{5}{c@{}}{$\bolds{\omega}$}\\[-6pt]
&&\multicolumn{5}{c@{}}{\hrulefill}\\
& $\bolds{\kappa}$ & \multicolumn{1}{c}{\textbf{1}} & \multicolumn{1}{c}{\textbf{2}} & \multicolumn{1}{c}{\textbf{4}}
& \multicolumn{1}{c}{\textbf{6}} & \multicolumn{1}{c@{}}{$\bolds{\omega^{*}}$}\\
\hline
{$\mathrm{FP}$}
& 1 & 2.99 (0.03) & 3.44 (0.05) & 3.90 (0.07) & 4.40 (0.08) & 4.87
(0.11) \\
& 4 & 5.53 (0.09) & 5.72 (0.10) & 6.18 (0.12) & 6.70 (0.14) & 7.18
(0.17)\\
& 8 & 6.77 (0.13) & 6.92 (0.15) & 7.33 (0.17) & 7.80 (0.20) & 8.26
(0.21)\\
& $\kappa^{*}$ & 7.50 (0.18) & 7.67 (0.20) & 8.08 (0.22) & 8.52 (0.24)
& 8.97 (0.25)\\[3pt]
{$\mathrm{FMP}_{H}$}
& 1 & 2.88 (0.04) & 3.17 (0.05) & 3.42 (0.07) & 3.58 (0.08) & 3.61
(0.09)\\
& 4 & 4.98 (0.11) & 5.14 (0.12) & 5.44 (0.15) & 5.65 (0.17) & 5.77
(0.18)\\
& 8 & 6.00 (0.15) & 6.12 (0.16) & 6.42 (0.19) & 6.66 (0.22) & 6.89
(0.24)\\
& $\kappa^{*}$ & 7.18 (0.27) & 7.47 (0.33) & 8.02 (0.50)& 8.41 (0.57)
& 8.80 (0.53)\\[3pt]
{$\mathrm{FMP}_{S}$}
& 1 & 2.80 (0.04)& 3.09 (0.05)& 3.30 (0.07)& 3.49 (0.07)& 3.49 (0.07)\\
& 4 & 4.88 (0.09)& 4.91 (0.10)& 5.26 (0.11)& 5.44 (0.13)& 5.42 (0.13)\\
& 8 & 5.90 (0.14)& 5.99 (0.14)& 6.34 (0.16)& 6.38 (0.17)& 6.59 (0.18)\\
& $\kappa^{*}$ & 6.90 (0.19)& 7.07 (0.19)& 7.25 (0.22)& 7.58 (0.23)&
7.86 (0.25)\\[3pt]
{$\mathrm{FMP}_{S}^{*}$}
& 1 & 2.60 (0.03) & 2.81 (0.04)& 3.03 (0.05)& 3.13 (0.06)& 3.13
(0.06)\\
& 4 & 4.11 (0.06) & 4.18 (0.07)& 4.40 (0.09)& 4.48 (0.10)& 4.46
(0.11)\\
& 8 & 4.57 (0.09)& 4.59 (0.09)& 4.74 (0.11)& 4.78 (0.12)& 4.73 (0.13)\\
& $\kappa^{*}$ & 4.58 (0.10)& 4.57 (0.10)& 4.65 (0.12)& 4.68 (0.13)&
4.62 (0.13)\\
\hline
\end{tabular*}
\end{table}

The average clustering error rates are 6.48\% (with standard error
1.28\%), 1.68\% (0.45\%) and 8.39\% (2.01\%) for Clusters 1--3 based on
the 100 simulated training data sets.
The accuracy of classification for the future trajectory to be
predicted for a partially observed trajectory in the test data depends
on the values $\tau$, the ``current'' time observed thus far. The
average classification error rate decreases with $\tau$, ranging from
27.5\% at 8:00 to 7.7\% at 20:00, implying that prediction accuracy
increases with $\tau$. Additional details regarding accuracy of
clustering and classification are compiled in Supplementary Material D [\citet{chi12}].

The prediction performances based on the proposed functional mixture
prediction (FMP) approach are summarized in\vadjust{\goodbreak} Table~\ref
{tabaucflm3sim}. The method $\mathrm{FMP}_{S}^{*}$ is the same as
$\mathrm{FMP}_{S}$, apart from that $\mathrm{FMP}_{S}^{*}$ assumes the
cluster/classifica\-tion memberships are known, serving as the gold
standard for prediction performance comparisons.
The results clearly demonstrate that $\mathrm{FMP}_{S}$ outperforms
$\mathrm
{FP}$ and $\mathrm{FMP}_{H}$, with relatively smaller values of TMIPE and
the associated standard errors indicating that $\mathrm{FMP}_{S}$ has
better prediction accuracy and is quite robust.
In $\mathrm{FMP}_{S}$, the prediction errors using $\omega=1$ and
$\omega
=2$ are close to each other and perform the best under various values
of $\kappa$. The optimal selection of $\omega$ appears to be different
from those obtained from our real traffic flow data. Although the
generated data based on the model estimates may reach a high level of
realism to traffic flow data, they may not be able to capture the
entire data features such as outlying curves that could influence the
prediction performance. In addition, classification errors of the
partially observed trajectory may also play a role in prediction.
We also compare the FMP method with the FPCP approach.
The simulation results are listed
in Supplementary Material D [\citet{chi12}]. The results suggest larger values of
$\omega$ for minimal prediction errors.
The intuition behind these results is that FPCP
treats the partial trajectory to be predicted as missing values,
especially when the data are more homogeneous within clusters and
contain less outlying curves as in the simulated data. Overall, the
results demonstrate that the proposed FMP approach outperforms the FPCP
approach.

%s7 #&#
\section{Concluding remarks and discussions}\label{sec7}

This study presents a methodological framework for uncovering traffic
flow patterns and predicting traffic flow. The proposed functional data
approaches, including classification and prediction, identify clusters
with similar traffic flow patterns, facilitating accurate prediction of
daily traffic flow. Although motivated by the subject of traffic flow
prediction, the proposed methodology is generally applicable and
transferable to the analysis and prediction of any longitudinally
collected functional data, such as city electricity usage or
degradation studies in manufacturing systems. The empirical results
demonstrate that our proposed method, functional mixture prediction,
which combine functional prediction with probabilistic functional
classification, can work reasonably well to predict traffic flow. We
conclude that taking traffic flow patterns into account can greatly
improve prediction performance as long as the traffic flow patterns can
be satisfactorily identified.

In the literature of intelligent transportation systems, conditional
expectation is commonly used as the measure of traffic flow
prediction/forecast of a future trajectory at a future time point or
short period. However, it may be interesting to consider probabilistic
forecasts [Gneiting (\citeyear{Gne08})], which take the form of probability
distributions over future trajectories.
A probabilistic forecast may engender a new way of thinking about
traffic flow prediction, which may give a better account of uncertainty
in potential flow trajectories. In this study,\vadjust{\goodbreak} our focus was on
predicting a future trajectory in the form of conditional expectation
for an up-to-date and partially observed trajectory. Under the
functional mixture prediction framework, a mixture of predictive
distributions of the potential trajectories could instead serve as an
ensemble for probabilistic forecasting. However, substantial efforts
would be needed to accomplish the goal of probabilistic forecasting for
traffic flow trajectories.

In addition to predictive accuracy, the real-time feature of traffic
flow information is important in traffic management. Given that the
components of our proposed model are estimated based on historical
data, as in the training data, the proposed method also serves as a
real-time prediction approach for predicting the future unobserved
traffic flow trajectory for a partially observed flow trajectory. The
fact that real-time information is quickly and easily updated will
facilitate the establishment of effective reporting systems for traffic
flow prediction. Furthermore, this article discussed single-detector
traffic prediction, a category crucial in supporting demand forecasting
as required in practice by operational network models. Future research
might extend to multiple-detector traffic prediction and will be
important in working toward the goal of better road network management.

\section*{Acknowledgments}
The author would like to thank Mr. Y.-C.
Zhang for computational assistance and Dr. J.~A.~D. Aston for carefully
reading the manuscript. The author would also like to express his
gratitude to the Editor, Associate Editor and referees, whose comments
and suggestions helped improve the paper.

\begin{supplement}[id=suppA]
\stitle{Supplement to: ``Dynamical functional prediction and classification,
with application to traffic flow prediction.'' by J.-M. Chiou\\}
\slink[doi]{10.1214/12-AOAS595SUPP} %[doi,text={...}] - jei reikia
%suskaldyti doi
\slink[url]{http://lib.stat.cmu.edu/aoas/595/supplement.pdf}
\sdatatype{.pdf}
\sdescription{This supplement contains a PDF (AOAS595SUPP.pdf) which
is divided into
four sections.
\textit{Supplement A}:
Selection of the number of clusters;
\textit{Supplement B}:
Bootstrap prediction intervals;
\textit{Supplement C}:
Additional results for traffic flow prediction;
\textit{Supplement D}:
Additional simulation details and results.}
\end{supplement}

% imsref loaded by akundreckaite, 2012-10-23 10:30:50
%
% imsref loaded by akundreckaite, 2012-10-23 12:52:50

\printaddresses

\end{document}